\documentclass[12pt, a4paper]{article}
\usepackage[font=footnotesize]{caption}
\usepackage[utf8]{inputenc}
\usepackage{graphicx}
\usepackage{authblk}
\usepackage{booktabs}
\usepackage{verbatim}
\usepackage{url}
\usepackage{caption}
\usepackage{subcaption}
\usepackage{indentfirst}
\usepackage{url}
\usepackage{amssymb}
\usepackage{amsmath}
\title{Node Accessibility Characterization \\ of Radially-Grown Structures}
\author{Alexandre Benatti$^1$, Roberto M. Cesar Jr.$^1$, and \\ Luciano da F. Costa$^2$}

\affil{
$^1$Institute of Mathematics and Statistics - DCC \\
University of S\~ao Paulo \\
Rua do Mat\~ao, 1010, \\ S\~ao Paulo, SP 05508-090 Brazil 
\\ \vspace{0.5cm}
$^2$S\~ao Carlos Institute of Physics - DFCM \\
University of S\~ao Paulo \\
Av. Trabalhador S\~ao-carlense, 400, \\ S\~ao Carlos, SP 13566-590 Brazil
}

\date{\today}

\begin{document}

\maketitle

\begin{abstract}
Complex systems have motivated continuing interest from the scientific community, leading to new concepts and methods. Growing systems represent a case of particular interest, as their topological, geometrical, and also dynamical properties change along time, as new elements are incorporated into the existing structure. In the present work, an approach is the case in which systems grown radially around some straight axis of reference, such as particle deposition on electrodes, or urban expansion along avenues, roads, coastline, or rivers, among several other possibilities. More specifically, we aim at characterizing the topological properties of simulated growing structures, which are represented as graphs, in terms of a measurement corresponding to the accessibility of each involved node. The incorporation of new elements (nodes and links) is performed preferentially to the angular orientation respectively to the reference axis. Several interesting results are reported, including the tendency of structures grown preferentially to the orientation normal to the axis to have smaller accessibility.
\end{abstract}

\section{Introduction}\label{sec:introduction}

As implied in their name, \emph{complex systems} have proven to be particularly challenging, motivating new concepts and methods, including network science (e.g.~\cite{barabasi2013network,newman2018networks,costa2007characterization,costa2011analyzing}). Among the several types of complex systems, we have growing structures in which basic elements are progressively incorporated into them. Examples include particle deposition (e.g.~\cite{elimelech2013particle,burd2009particle}) and urban expansion (e.g.~\cite{robson2012urban,ding2019urban,van2021impact}), among many other interesting situations. These types of complex systems are of particular interest because their geometry, topology, and dynamics typically change as they grow, motivating several related investigations.

The present work aims at investigating the topological organization, from the specific point of view of \emph{accessibility} (e.g.~\cite{travenccolo2008accessibility, travenccolo2009border, benatti2022complex}) of structures growth around a straight reference axis, which is henceforth referred to as \emph{radial growth}. This type of system includes particle deposition onto surfaces or electrodes (e.g.~\cite{guelcher2000aggregation,dabros1983direct}), urban expansion along avenues, roads, or coastline (e.g.~\cite{robson2012urban,ding2019urban}). This type of growing structure has special interest because the resulting system will depend not only on the aggregation rules but also on the shape of the reference axis, which acts as a guideline along the development.

Because the reference straight axis establishes a well-defined orientation, it is of particular interest to perform the growth while taking into account the relative orientation of the aggregation respectively to that of the axis. This issue has special importance in several real-world cases. For instance, in the case of urban expansion around an avenue, new street segments can be incorporated at an angle that is either parallel or normal to the straight reference, depending on whether the new blocks are expected to be located close (e.g.~in the case of resorts developing along the coastline) or further away from the reference axis (e.g.~in the case of residential boroughs in large cities).

For the sake of generality, the growing structures are represented here as graphs (or complex networks). Furthermore, these graphs are assumed to grow by incorporating new edges and nodes chosen from an underlying lattice. The latter is adopted in order to provide some intrinsic organizational regularity at the smallest topological scales, as is often the case of growing structures such as urban regions (typically orthogonal) and biological tissues (also nearly regular). The incorporation of the orthogonal lattice constraint also contributes to having more definite parallel or orthogonal preferential attachment respectively to the orientation of the axis. Henceforth, the probabilities of adding normal or parallel links to one of the nodes of the growing system are represented as the parameters $p_n$ and $p_p$, respectively, with $p_p=1-p_n$.

The initial state of the system corresponds to a linear sequence of nodes, onto which new links (constrained by the underlying lattice) are successively incorporated preferentially to the relative orientation between the link and the reference axis. A recent model of radial growth~\cite{benatti2023modeling}, which incorporates the above characteristics, is henceforth adopted in this work as a means to obtain several instances of the growing structures.

The main purpose of the present work consists of studying the topological properties, more specifically the \emph{node accessibility} (e.g.~\cite{travenccolo2008accessibility,travenccolo2009border, benatti2022complex}) of structures growth under distinct configurations. We will be mainly interested in structures with links incorporated preferentially to their relative orientation respectively to the reference axis. Emphasis has been placed on the node accessibility measurement since it provides a multi-scale indication of the structural relationship between a node and its successive neighbors. Informally speaking, the accessibility of a specific node at topological scale $h$ (the topological distance to the considered neighbors) expresses how effectively that node can potentially interact (e.g.~during a traditional random walk) with those neighboring nodes.

Basically, given a specific node in a network, we estimate its transition probabilities to nodes at successive hierarchical neighborhoods, as implied by traditional uniform random walks. The exponential entropy (e.g.~\cite{campbell1966exponential,pielou1966shannon}) of these probabilities are then calculated and used for respective characterization of the number of nodes at a given hierarchy $h$ that are effectively accessible from the reference node. It can be shown that the accessibility at a given hierarchical neighborhood varies from 0 to the number of neighbors contained in that neighborhood. As such, the node accessibility naturally provides a measurement that can be understood as a generalization of the node degree (e.g.~\cite{benatti2021accessibility,benatti2022complex}) not only along successive hierarchies but also by possibly taking non-integer values. The node accessibility concept has been employed for the study of several abstract and real-world systems (e.g.~\cite{vera2022neumann,lawyer2015understanding,de2014role,amancio2015comparing,amancio2015complex,tohalino2018extractive,tokuda2022impact}).

Markedly distinct types of structures are obtained by varying the parameter $p_n$, which have been found to have diverse respective fractionary dimensions~\cite{benatti2023modeling}. A study of random walks performed on these different types of structures has also been reported in~\cite{benatti2023modeling}, taking into account the marginal density probabilities obtained along the straight reference axis and normally to it.

The experiments addressed in the present work refer to structures obtained along varying development stages and consider several preferential probabilities and hierarchical extensions. In particular, we consider three main situations characterized by anysotropic growth favoring normal or parallel connections, as well as a third situation characterized by isotropic connections ($p_n=p_p$). Several interesting results are described and discussed, especially the tendency of anysotropic growth favoring parallel connections yielding enhanced node accessibility when compared to normal preferential connections (considering the same growth stages). This suggests that growing systems that prioritize depth instead of breadthwould tend to be characterized by relatively smaller node accessibilities.

The present work starts by presenting and illustrating the main concepts and methods and then proceeds by describing the experimental approach, which is followed by a presentation and discussion of the respectively obtained results.

\section{Concepts and Methods}\label{sec:methods}

This section describes the main concepts and methods used in this work, covering node accessibility and modeling radial growth.

\subsection{Node Accessibility}

Given a set of probabilities $p_i$, $i=1, 2, \ldots, N$, so that their sum is equal to one, the respective \emph{entropy} can be expressed as:
\begin{align}
   \varepsilon = - \sum_{i=1}^{N} p_i \, \log(p_i)
\end{align}

The respective \emph{exponential entropy} (e.g.~\cite{jost2006entropy}) can then be obtained as:
\begin{align}
   \alpha = e^{\varepsilon}
\end{align}

It can be verified that this quantity is necessarily contained in the interval $[0,N]$, with the maximum value $N$ being obtained when all probabilities are identical to $p_i=1/N$.

The exponential entropy has been employed as a means of generalizing (e.g.~\cite{campbell1966exponential,pielou1966shannon}) the concept of node degree in graphs and complex networks. First, a traditional random walk (or any other type of dynamics of particular interest) is performed on the network, therefore providing an estimation of the transition probabilities between any pair of nodes. Given a network and one of its nodes $i$, the respective neighbors at each successive hierarchical neighborhood $h$ are identified, corresponding to the network nodes that are at distance $h$ from the node of interest. The transition probabilities from the reference node to any of this set of neighbors at a fixed hierarchy are necessarily normalized, allowing the respective entropy to be estimated as above.

\begin{figure}
  \centering
     \includegraphics[width=.9 \textwidth]{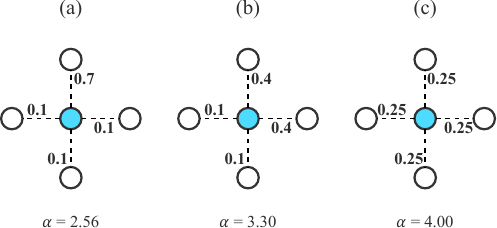}
   \caption{Illustration of the node accessibility in quantifying how effectively the reference node (in cyan) can access the neighbors at the $h-$th neighborhood. Each dashed link represents a path from the reference node up to its $h-$th neighbors. The numbers indicate the respective transition probabilities. The maximum accessibility is equal to the number of neighbors at the considered hierarchy, which takes place when all transition probabilities are identical, as illustrated in (c).}\label{fig:acc_ex}
\end{figure}

Figure~\ref{fig:acc_ex} illustrates the potential of the node accessibility~\cite{travenccolo2008accessibility} in quantifying the interaction between a reference node and its neighbors at the $h$-th hierarchy, which are reachable through varying respective /transition probabilities.

Observe that, for non-weighted complex networks, the node accessibility becomes identical to the node degree for $h=1$, which is in agreement with the former measurement being understood as a generalization of the latter.

The node accessibility has been found to be closely related to the \emph{borders} of graphs and complex networks (e.g.~\cite{travenccolo2009border}). More specifically, nodes that are at or close to the border of the respective network tend to present smaller node accessibility values than those nodes that are close to the core or center of the same network.

It is interesting to consider that, given a complex network representation of some specific structure (radially-grown regions in the case of the present work), it is not necessarily the case that the node accessibility should be generally increased throughout the network. This depends on the type of structure and dynamics being considered. For instance, in the case of urban mobility, high values of node accessibility tend to indicate that several nodes can be effectively accessed by random walks initiated at the reference node. This is, in principle, a property of interest in an urban region, because it would allow the respective resources to be more effectively accessed. However, there are situations in which low, instead of high, node accessibility could be preferred. For instance, in the case of disease propagation, low values of node accessibility could help minimize and/or delay the disease spreading.

The node accessibility values obtained for other types of systems will also have respective interpretations. For instance, in the case of particle deposition, the accessibility of a given node tends to quantify the contact interaction (related to the propagation of forces, heat, charge, vibrations, etc.) between that particle and other particles at successive neighborhoods.

It is also worth noticing that the overall node accessibility values obtained from a given radial structure do not reflect only the respective outer borders, but also small gaps or voids, which can be understood as inner borders. In this respect, the structures obtained tend to have the highest node accessibility values within their more central regions (e.g.~\cite{travenccolo2009border,viana2010characterizing}).

\subsection{Radial Growth Modeling Approach}\label{sec:Approach}

There are several ways in which a growing region can be initiated. For instance, it can start at a specific small region containing some important resource, and then grow in an approximately uniform way around this initial region. Another possibility is that the structure is initiated as a prolongation of another already existing reference structure to which it will remain adjacent. 

The present work focuses on a specific case of the latter possibility, in which the growing structure emanates from an existing straight reference axis, which can be a street, avenue, road, or highway in the case of urban regions, or a plane surface or straight electrode in the case of particle deposition. More specifically, new edges (assumed to be non-directed) are progressively incorporated starting from this reference, which is assumed to be straight. The growing region is limited to an extent $L$ along the reference axis.

In the present work, we adopt the methodology of growing radial structures described recently in~\cite{benatti2023modeling}. As in that work, each new link, which has the length determined by the size of the cells of the underlying orthogonal lattice, which has been adopted in order to ensure structural regularity at the smallest topological scales while emphasizing the possibility of new links having orientation parallel or normal to the reference axis.

At each time step $e$, one of these possible new links is incorporated to some \emph{adjacent, already existing node}. The probability of choosing the orientation of a new link is preferential to its angular orientation respectively to the reference axis, which can be either parallel or normal. The parameter $p_p$ henceforth controls the preference to parallel connections, with the normal probability being $p_n=1-p_p$.

\begin{figure}
  \centering
     \includegraphics[width=.99 \textwidth]{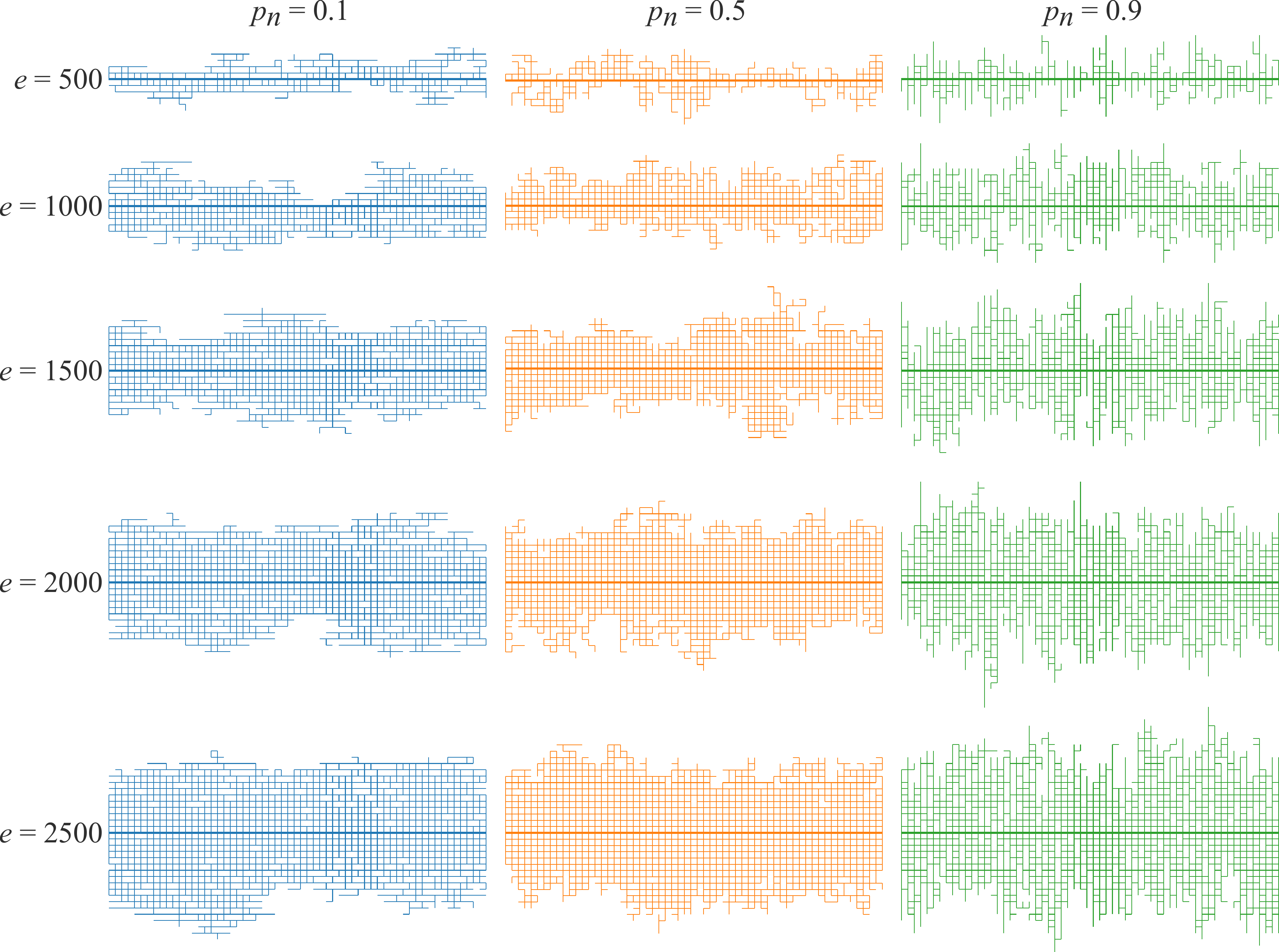}
   \caption{Examples of synthetic regions obtained along growth stages $e$ = 500, 1000, 1500, 2000, and $2500$ for $p_=0.1, 0.5$ and $0.9$ and $L=60$.}\label{fig:mapEx}
\end{figure}

Figure~\ref{fig:mapEx} illustrates stages $e=500, 1000,1500,2000,2500$ along the development stages $e$ of a synthetic region with $L=60$ and $p_n=0.1,0.5,0.9$. The value of $e$ of a synthesized network corresponds to its current number of edges. The value of $H$ has been chosen large enough in order that the growing structure does not touch the horizontal borders of the enclosing rectangle.

\section{Experimental Approach}

The experiments performed in the present work involve radially growing structures with different parameters and quantifying their respective topological properties in terms of node accessibility measurements. We consider three basic situations: (i) growth preferential to the parallel orientation ($p_p > p_n$); (ii) isotropic growth ($p_p = p_n = 0.5$); and (iii) growth preferential to the normal orientation ($p_p < p_n$).

\begin{figure}
  \centering
     \includegraphics[width=.8 \textwidth]{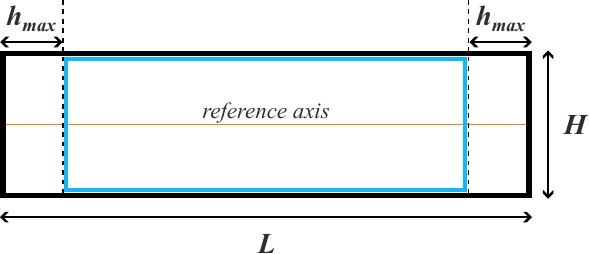}
   \caption{Diagram illustrating the region where the structure is allowed to grow, which has dimension $L \times H$. In order to \emph{avoid border effects}, the value $H$ is selected to be large enough, and the two lateral regions with extent $h_{max}$ (the maximum considered neighborhood hierarchy) are excluded from the analysis. Only the node accessibility values of nodes comprised in the region delimited by the blue rectangle are considered for the generation of the node accessibility density.}\label{fig:reference_axi}
\end{figure}

In all cases, the straight reference axis consisted of a chain of L nodes. The growth can take place within the rectangle of size $L \times H$. Also, as illustrated in Figure~\ref{fig:reference_axi}, two buffer zones of extension $h_max$ (the maximum hierarchical level considered in each analysis) are considered at both extremities of the region allocated for growth. These regions are implemented in order to minimize the influence of the network borders on the respective measurements~\cite{viana2010characterizing}.

The structures are grown by using the methodology previously described in ~\cite{benatti2023modeling} (see also Section~\ref{sec:Approach}), and the node accessibility is then calculated for instances of the structures along their growing steps, which corresponds to the number $e$ of links added since the beginning of the growth.

\section{Results and Discussion}\label{sec:results}

This section presents the main experimental results and respective discussion. We start by presenting the distribution of node accessibility mapped onto some of the synthesized structures and then present the node accessibility densities obtained for the several considered parametric configurations. In all cases, we have $L=60$, $H=50$, and $h_{max}=10$. The chosen value of $H$ is large enough to avoid the growing regions to touch the horizontal borders of the enclosing region.

\begin{figure}
  \centering
    \includegraphics[width=1 \textwidth]{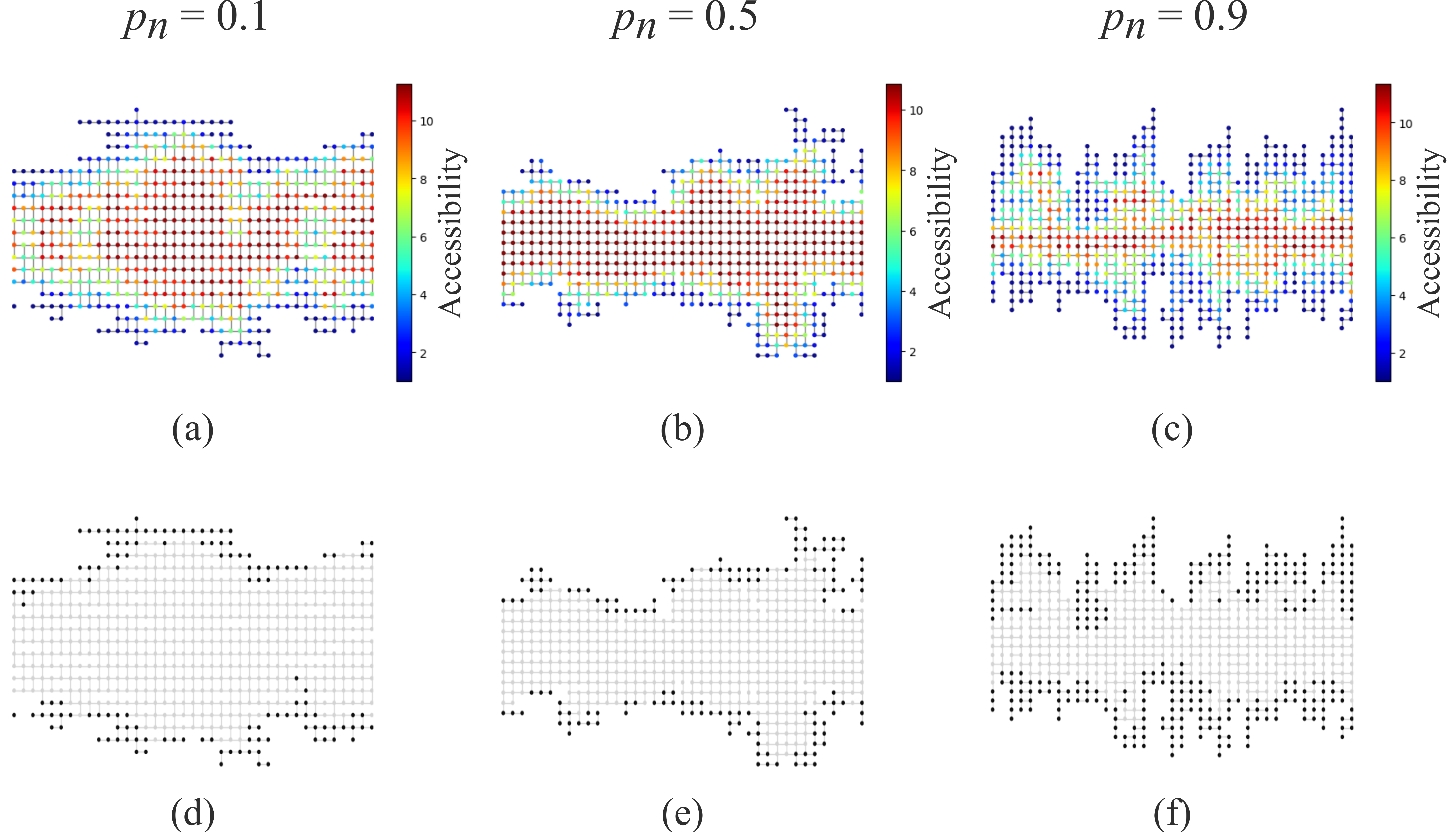}
   \caption{Examples of regions obtained for $p_n=0.1, 0.5, 0.9$ for $e=1500$ and $h=3$, with the node accessibility values shown in terms of a respective heatmap (a--c). The respectively obtained border nodes (respectively shown in d--f) correspond to the nodes with accessibility smaller or equal to $6$. }\label{fig:borders}
\end{figure}

Figure~\ref{fig:borders}(a--c) depicts examples of regions obtained respectively to $p_n=0.1, 0.5, 0.9$ for $e=1500$ and $h=3$. The largest node accessibility values can be observed around the reference axis, decreasing steadily at nodes located further away. The structure grown preferentially to the normal (c) is characterized by the narrowest region of relatively large values of node accessibility. Also shown (d--f), are the borders of the respective structures, marked (in black) as the nodes with accessibility smaller than $6$. These results indicate that wider/longer borders were obtained for normal growth (f).

\begin{figure}
  \centering
     \includegraphics[width=.99 \textwidth]{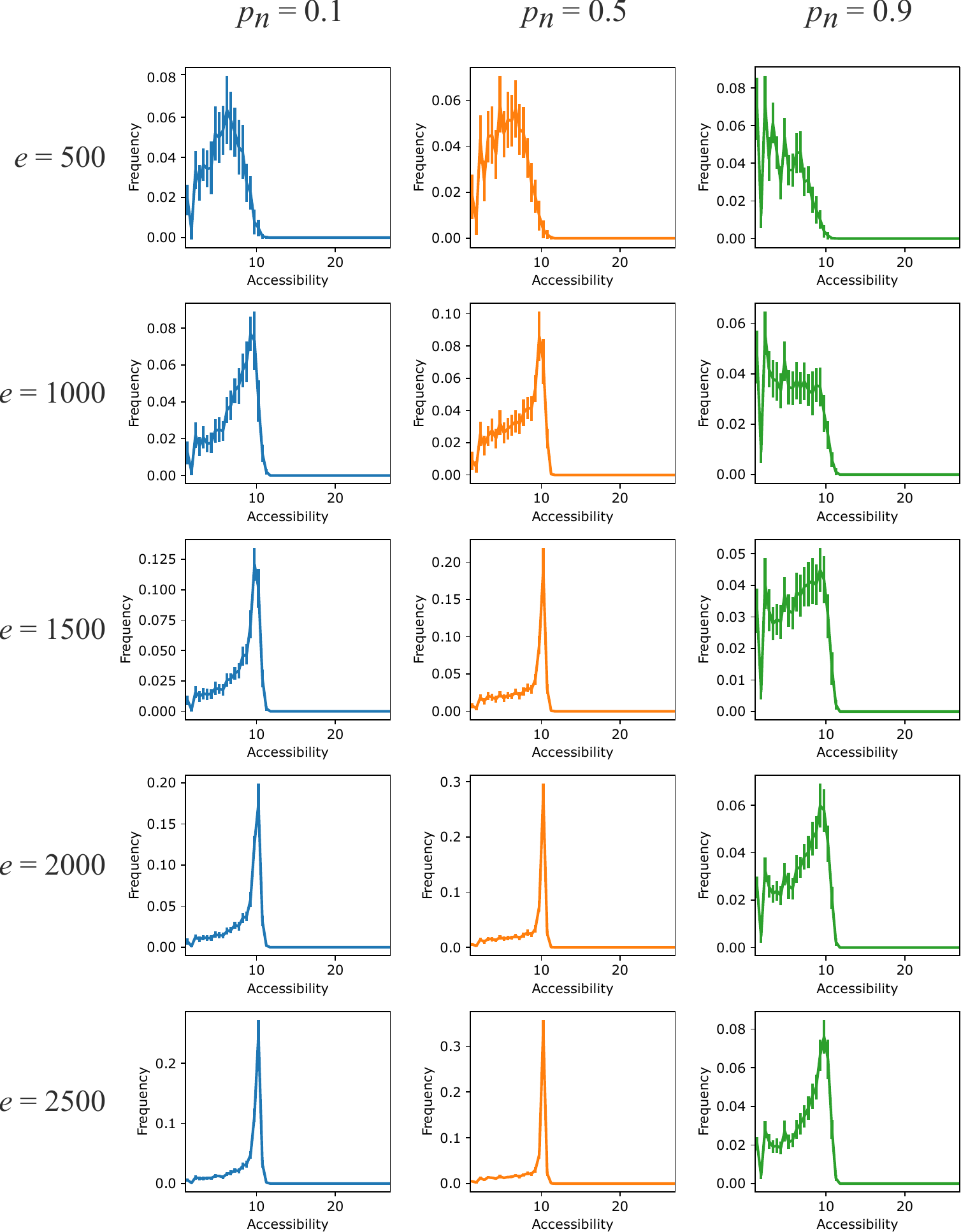}
   \caption{The distribution (average $\pm$ standard deviation) of the node accessibility respective to synthetic regions obtained at $e=500, 1000, 1500,$ and $2500$ for $p_n=0.1, 0.5, 0.9$ and $h=3$. The average node accessibility tends to increase with $e$.}\label{fig:hist_h3}
\end{figure}

\begin{figure}
  \centering
     \includegraphics[width=.99 \textwidth]{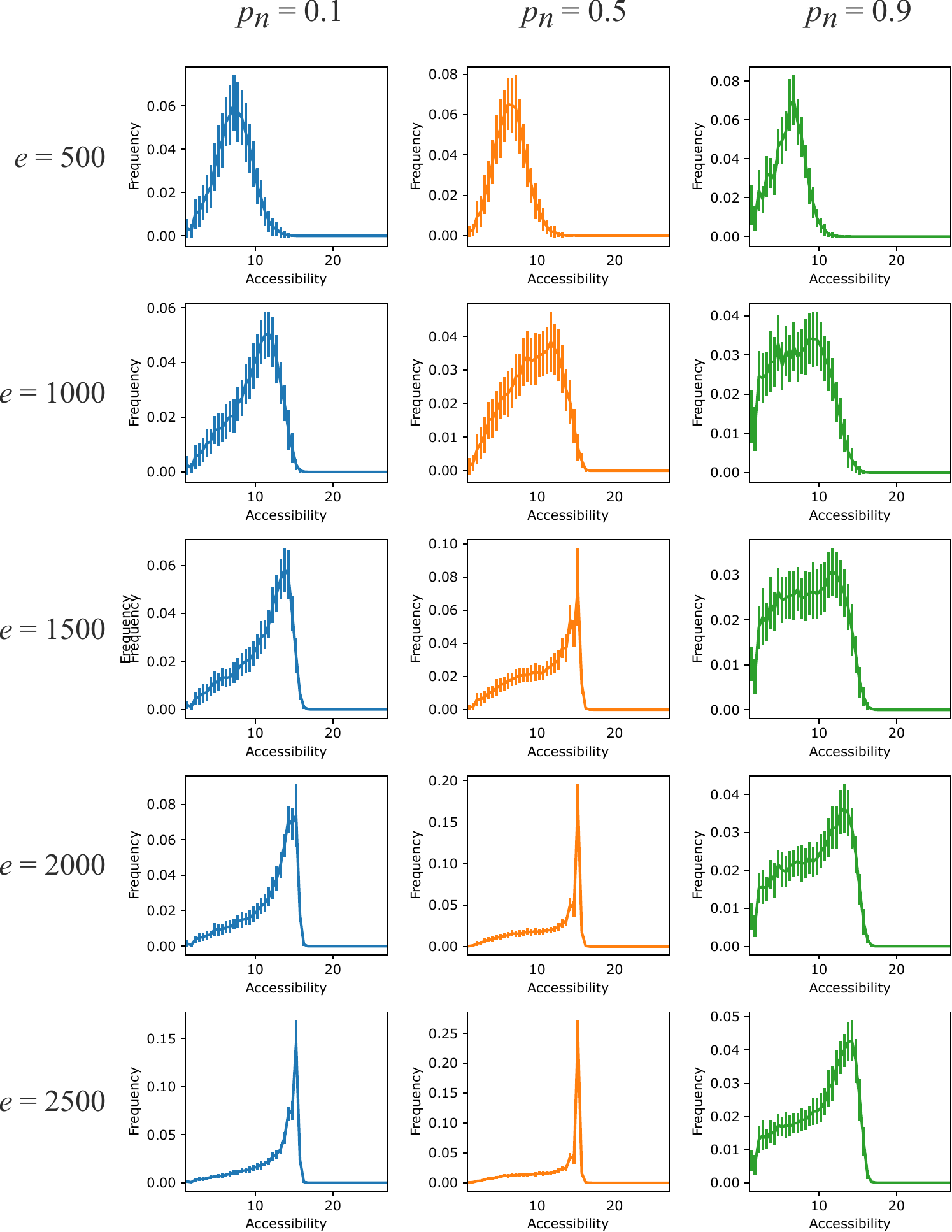}
   \caption{The distribution (average $\pm$ standard deviation) of the node accessibility respective to synthetic regions obtained at $e=500, 1000, 1500,$ and $2500$ for $p_n=0.1, 0.5, 0.9$ and $h=5$.}\label{fig:hist_h5}
\end{figure}

\begin{figure}
  \centering
     \includegraphics[width=.99 \textwidth]{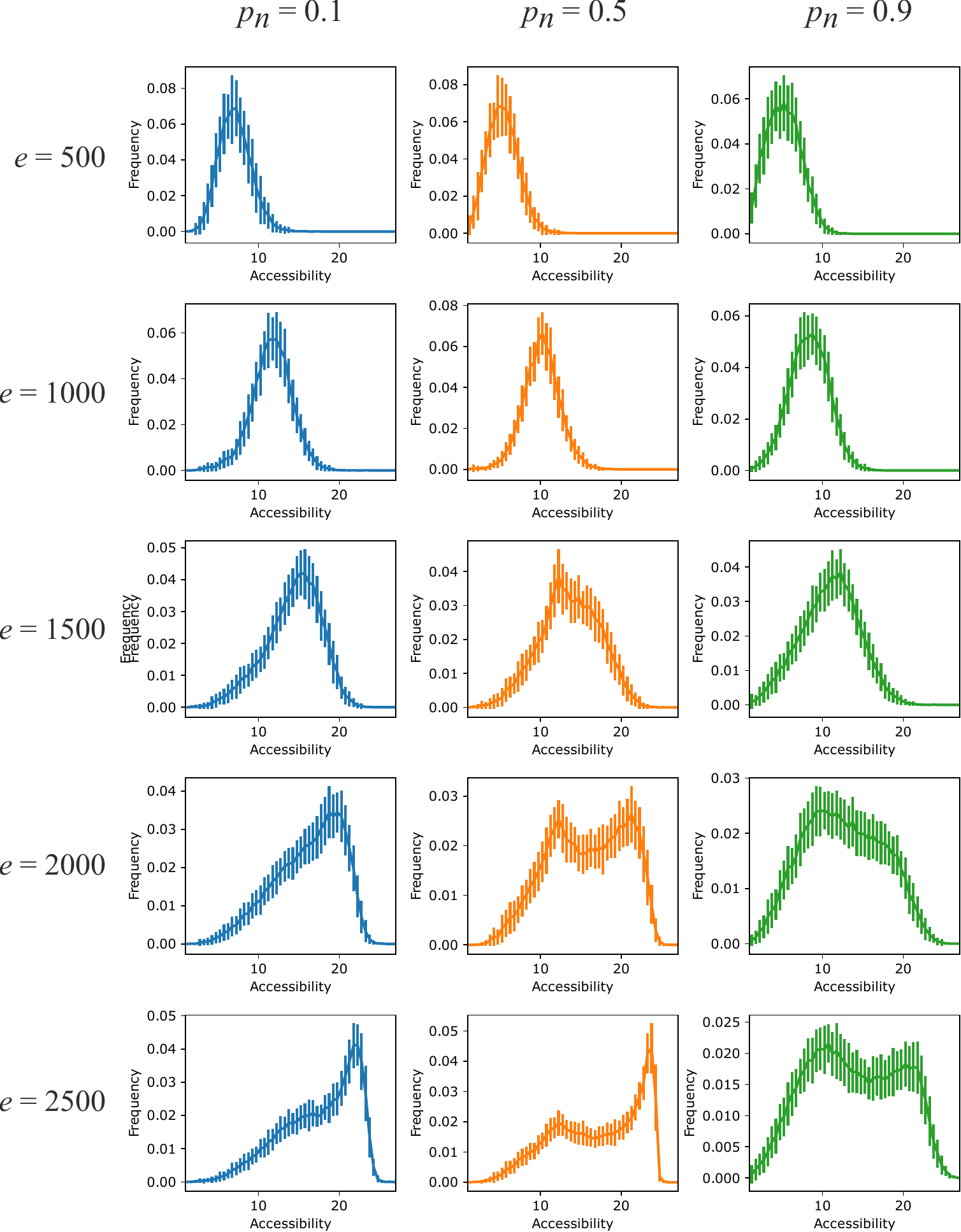}
   \caption{The distribution (average $\pm$ standard deviation) of the node accessibility respective to synthetic regions obtained at $e=500, 1000, 1500,$ and $2500$ for $p_n=0.1, 0.5, 0.9$ and $h=10$. }\label{fig:hist_h10}
\end{figure}

The densities of node accessibility obtained for the parametric configurations $L=60$, $p_n = {0.1,0.5,0.9}$ are presented respectively to $h=3, 5$ and $10$ in Figures~\ref{fig:hist_h3},~\ref{fig:hist_h5}, and~\ref{fig:hist_h10}.

Several interesting results can be observed from Figures~\ref{fig:hist_h3},~\ref{fig:hist_h5}, and~\ref{fig:hist_h10}. First, we have that the node accessibility values tend to increase with $h$ in all cases. That could be expected because more neighboring nodes tend to be encountered at longer topological distances from the reference node. The node accessibility values also tend to increase, up to eventual saturation, with the values of $e$. The dispersion of the node accessibility around its peak also tends to decrease steadily near the saturation. Observe that more links need to be added before saturation is observed in the case of larger values of $h$.

The isotropic configuration $p_n=p_p=0.5$ (in orange) does not necessarily lead to maximum overall accessibility, yielding a number of border nodes that is often smaller than those obtained for other non-isotropic configurations. Indeed, the configurations $p_n<0.5$ (in blue) tended to yield larger values of node accessibility. At the same time, configurations favoring normal growth ($p_n>p_p$) (in green) led to particularly small values of node accessibility, with relatively large border regions.

\begin{figure}
  \centering
    \includegraphics[width=.7 \textwidth]{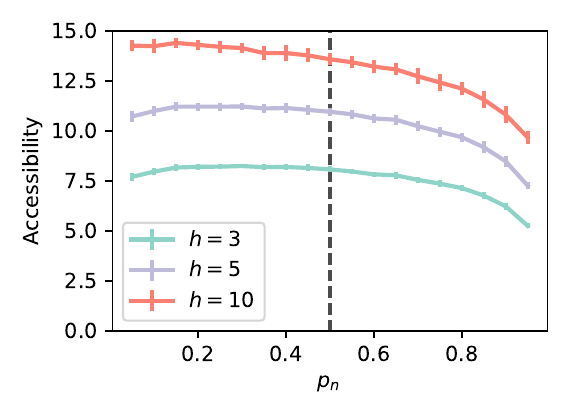}
   \caption{The mean $\pm$ standard deviation of the node accessibility values obtained for sets of 30 regions synthesized for $p_n = 0.05, 0.1, 0.15, \ldots, 0.95$ while considering $e=1500$ and $h=3, 5, 10$. The dashed line indicates the isotropic configuration $p_n=0.5$.}\label{fig:asym}
\end{figure}

Figure~\ref{fig:asym} presents the mean $\pm$ standard deviation of the mean node accessibility in terms of $p_n = 0.05, 0.1, 0.15, \ldots, 0.95$ respectively to synthetic regions obtained for $e=1500$ and $h=3, 5, 10$.

The configurations adopting small values of $p_n$, especially those obtained for $h=5$ and $10$, resulted in substantially higher node accessibility values than those obtained for the higher values of $p_n$ or even for $p_n=0.5$. This is related to the tendency of the respective regions to present fewer border nodes, which have smaller node accessibility, than the other configurations. This tendency is further illustrated in the first row of Figure~\ref{fig:borders} respectively to $e=1500$ and $h=3$, which presents examples of the regions obtained for $p_n=0.1, 0.5, 0.9$.

\section{Concluding Remarks}

Complex structures and systems have attracted increasing interest from the scientific and technological communities. Growing structures, in particular, have defined a challenging research subject due to the various interacting aspects involved. This has led to continued and growing interest.

The present work reported a study of the node accessibility properties of synthetically generated regions (networks) grown around a straight axis according to varying orientation-specific growth probabilities. The distribution of node accessibility was then estimated for each obtained synthetic region, with several interesting results.

Among the several interesting results obtained, we have that structures grown preferentially with angular orientation parallel to the reference axis tended to yield particularly high values of node accessibility which are, in several cases, larger than those obtained for the isotropic growth ($p_n=p_p=0.5$). Also of special interest is the observation that the node accessibility values tend to undergo a relatively abrupt decrease with large values of $p_n$, resulting in networks with large borders.

Future related works include the consideration of more general types of references, such as rings, as well as embedding spaces with dimension larger than 2 (which has been adopted in the present work). It would also be interesting to consider the attachment of new links to be preferential not only to the relative orientation with the reference axis, but also to take into account the local density of connections as well as the distance from the adopted reference structure.

\section*{Acknowledgments}
The authors are grateful to MCTI PPI-SOFTEX (TIC 13 DOU 01245.010\\222/2022-44) and FAPESP (grant 2022/15304-4). Roberto M. Cesar Jr. thanks CNPq. Luciano da F. Costa thanks CNPq (grant no.~307085/2018-0).

\bibliography{ref}
\bibliographystyle{unsrt}

\end{document}